\documentstyle[12pt,psfig]{article}

\def\ba{\begin{array}}
\def\bea{\begin{eqnarray}}
\def\be{\begin{equation}}
\def\cs{c\!\!/ }
\def\dag{\dagger}

\def\ea{\end{array}}
\def\eea{\end{eqnarray}}
\def\ee{\end{equation}}
\def\eh{\hat e}
\def\eps{\epsilon}
\def\fpic{f_{\pi}^2}
\def\fr{\frac}
\def\Lam{\Lambda}

\def\lra{\leftrightarrow}
\def\l{\left}
\def\nn{\nonumber}
\def\om{\omega}

\def\rh{\hat r}
\def\r{\right}

\def\sig{\sigma}

\def\ss{s\!\!\!/ }

\def\Tr{{\,\mbox{Tr}}}
\def\veps{\varepsilon}

\def\bel{\begin{flushleft}}
\def\eel{\end{flushleft}}
\def\bit{\begin{itemize}}
\def\eit{\end{itemize}}

\def\fpis{{f_\pi^2}}
\def\tr{\mbox{Tr}}
\def\rarr{\rightarrow}

\def\eh{\hat e}
\begin{document}
\begin{titlepage}
\begin{center}
\vspace*{2.0cm}
{\Large\bf The hyperon--nucleon  interaction potential in the \\ 
bound state soliton model: the $\Lambda N$ case }
\vskip 1.5cm

{G.L. Thomas$^a$, V.E. Herscovitz$^a$ , C.L. Schat$^b$
and N.N. Scoccola$^{c,}$\footnote{Fellow
of the CONICET, Argentina.} }
\vskip .2cm
{\it
$^a$ Instituto de F\'{\i}sica, Universidade Federal do Rio Grande do Sul, \\
Av. Bento Gon\c calves 9500, 91501-970, C.P. 15051,  Porto Alegre, RS, Brazil.  \\
$^b$ Centro Brasileiro de Pesquisas F\'{\i}sicas , DCP,  \\
Rua Dr. Xavier Sigaud 150, 22290-180, Rio de Janeiro, RJ, Brazil. \\
$^c$ Department of Physics, Comisi\'on Nacional de Energ\'{\i}a At\'omica,
          Av. Libertador 8250, \\
(1429) Buenos Aires, Argentina. }
\vskip 2.cm
November 1998 \\
\vskip 2.cm
{\bf ABSTRACT}\\
\begin{quotation}

We develop the formalism to study the hyperon-nucleon interaction 
potential within the bound state approach to 
the $SU(3)$ Skyrme model. The general framework is illustrated
by applying it to the diagonal $\Lambda N$ potential.  
The central, spin-spin 
and tensor components of this interaction are obtained 
and compared with those derived using alternative schemes.

\end{quotation}

\end{center}
\vspace{1cm}
state soliton model
 
\end{titlepage}

\section{Introduction}

The long standing interest in the hyperon-nucleon two-body interaction
is motivated by several reasons. For example, such interaction is the fundamental 
building block for a microscopic understanding of hypernuclei \cite{DG84}. In addition, 
the inclusion of the strangeness degrees of freedom calls for the extension of the models 
of the nucleon-nucleon potential so as to provide a unified coherent picture of the
baryon-baryon interaction. Nowadays quantum chromodynamics (QCD) is considered 
to be the fundamental theory for the strong
interactions. However, the low momentum- and energy-transfer region,
which is the relevant one for nuclear physics, is dominated by
non-perturbative processes. Consequently, a first principles evaluation of 
the baryon-baryon interaction in terms of quarks and gluons has not been 
possible up to now. In this situation one has to resort to models. 
In the phenomenological one boson exchange (OBE) model \cite{Ho80}  
the nucleon-nucleon interaction is described through the exchange of 
different mesons, supplemented by a short range  repulsion. In 
the case of the hyperon-baryon interaction, the Nijmegen 
\cite{NRS79} and J\"ulich \cite{HHS89} potentials are obtained by extending the OBE model to 
include the degrees of freedom that carry strangeness. In this way the quite limited
$\Lambda N$ and $\Sigma N $ scattering data can be described within a
unified picture at the price of introducing a rather important 
number of adjustable parameters. In fact, since the experimental data are 
sufficiently crude
they can be reproduced using various sets of parameters, out of which two representative 
examples are those which define the models D and F of 
Ref.\cite{NRS79}.
 
Here we will follow an alternative approach which is based on the Skyrme 
model \cite{Sk61,Za86}. This model relies on the fact that for a large 
number of colours $N_c$, QCD becomes equivalent to a
local field theory of mesons \cite{tH74} where the nucleons emerge as
chiral topological solitons \cite{Wi79} of the meson effective field theory.
The Skyrme model is the simplest choice of such a theory. It provides a 
reasonable good description of the $SU(2)$ baryon properties
and has already been implemented for the
construction of the nucleon-nucleon potential, as reviewed in
\cite{EK96,WW92}. To include strangeness in this scheme 
we will consider the bound state approach (BSA)\cite{CK85,SNNR88} 
extension of the Skyrme model to flavour $SU(3)$.  
In this way we complement the work of Refs.\cite{KE90,SSG95},
where the hyperon-nucleon potentials have been studied in the 
collective coordinates approach (CCA) to the $SU(3)$ 
Skyrme model. Differently from the CCA, where strangeness appears
as a collective rotational excitation, in the BSA hyperons are described 
as bound states of kaons in the background field of a $SU(2)$ soliton.
It is worthwhile to point out that these
soliton models have both the merit of describing the different
baryonic sectors ($B=1$ and $B=2$, $B$ being the baryon number) in a single
comprehensive framework. Moreover, the corresponding predictions are essentially 
parameter free since, in principle, all the parameters in the effective action 
can be fixed by the meson phenomenology in the $B=0$ sector. 

The present work constitutes a first step towards a general discussion of 
the hyperon--nucleon potential within the bound state model.  It is similar 
in spirit to previous $NN$ potential calculations done in the $SU(2)$ Skyrme 
model, although technically much more involved.  
The interaction potential can be written in the general form 
\bea
\label{pot}
V_{H N}(\vec r)  &=& V_C(r)  \ {\cal O}_C + V_{S}(r)  \ {\cal O}_{S} + V_T(r)  \ {\cal O}_T  
\eea
where 
${\cal O}_C = I_{2 \times 2}^H I_{2 \times 2}^N , 
\ {\cal O}_S = \vec\sigma_H \cdot \vec\sigma_N  , 
\ {\cal O}_T  =  
3 \ \vec \sigma_H \cdot \rh \ \vec \sigma_N \cdot \rh - \vec \sigma_H \cdot \vec \sigma_N $ 
and $V_C(r), \ V_S(r), \ V_T (r)$ are the central , spin-spin and tensor parts of the 
interaction, respectively. They can be decomposed into an isospin independent contribution 
$V^+$ and an isospin dependent one $V^-$, so that 
\bea
V_{C,S,T}(r) &=& V^+_{C,S,T}(r) + V^-_{C,S,T}(r) \ \vec \tau_H \cdot \vec \tau_N \ .
\eea
The contributions depending on angular momentum, like the
spin-orbit coupling, are not shown because they remain unaccessible
within the approximations used in our calculation. The new feature of
the $SU(3)$ case is the strangeness exchange interaction mediated by
kaons. In the language of the OBE models these interactions are of first
order in terms of kaon exchanges, but as well as the direct
contributions include higher orders ( more than one boson exchange) in
the $SU(2)$ sector. The general framework is illustrated by the explicit
calculation of the diagonal $\Lambda N$ potential, which is the simplest one due to
the absence of isospin interactions.  

The article is organized as follows. In Sec.2 we review briefly the
BSA and in Sec.3 we describe the general procedure to obtain the interaction
Lagrangian. In Sec.4 we show how to obtain the explicit form of the 
interaction potential in the $\Lambda N$ case. In Sec.5 we present our numerical
results. Finally, in Sec.6 our conclusions are given. 
Some useful expressions needed for the evaluation of the collective part 
of the matrix elements appearing in the potential can be found in the 
Appendix. 

\section{The bound state soliton model}

The bound state soliton model has been discussed in detail in the literature 
\cite{CK85,SNNR88}. Therefore, only a brief outline of its main features will be
presented here. 
The starting point is an effective $SU(3)$ chiral action which includes an explicit
symmetry breaking term. It has the form 
\be\label{lag}
\Gamma = \int d^4x \Big( {\cal L}_2 + {\cal L}_4 + {\cal L}_{\rm SB} \Big)
+ \Gamma_{\rm WZ} \ , 
\ee
where ${\cal L}_2 $ is the well-known non linear $\sigma$-model lagrangian density,
\bea
{\cal L}_2 &=& - {f^2_\pi \over 4} \Tr\l[L_\mu L^\mu \r] \ ,
\eea 
and ${\cal L}_4 $ is the Skyrme stabilizing term,  
\bea
{\cal L}_4 &=&  {1\over{32 \eps^2}} \Tr \Big[ [L_\mu, L_\nu] [L^\mu, L^\nu] \Big] \ ,
\eea
with the left current $L_\mu$  expressed in terms of the $SU(3)$ 
valued chiral field $U(x)$ as $L_\mu=U^\dagger \partial_\mu U$. 
Here $f_\pi$ is the pion decay constant and $\eps$ is the so-called Skyrme 
parameter.
 
The non-local Wess-Zumino action
 $\Gamma_{\rm WZ}$  is given by 
\bea
\Gamma_{\rm WZ} &=& - \fr{i N_c}{240 \pi^2} \int_{D5} d^5x \ 
\veps^{\mu\nu\alpha\beta\gamma} \Tr \l[ L_\mu L_\nu L_\alpha L_\beta L_\gamma \r] \ , 
\eea
where the domain of integration is a five dimensional disk $D_5$  whose boundary is 
space--time.  
The symmetry breaking term ${\cal L}_{\rm SB}$ takes into account the 
difference between the mass of the kaon $m_K$ and the mass of the pion $m_\pi$ 
as well as the difference between $f_\pi$ and the kaon decay constant $f_K$. 
It is given by 
\bea
{\cal L}_{\rm SB} & = &
 { f_\pi^2 m_\pi^2 + 2 f_K^2 m_K^2 \over{12} }
 \Tr \left[ U + U^\dagger - 2 \right] 
  + \sqrt{3}  { f_\pi^2 m_\pi^2 - f_K^2 m_K^2 \over{6} }
\Tr \left[ \lambda_8 \left( U + U^\dagger \right) \right] \  
\nonumber \\
&  & 
- { f_K^2 - f_\pi^2\over{12} }
\Tr \left[ (1- \sqrt{3} \lambda_8)
\left(
U L_\mu L^\mu + U^\dagger R_\mu R^\mu \right)
\right] \ ,   
\label{sb}
\eea
with $\lambda_8$ being  the eighth Gell-Mann matrix and  $R_\mu$  the right current  
$R_\mu=U \partial_\mu U^\dag$.

To describe the $B=1$ soliton sector we introduce the ansatz \cite{CK85}
\be   \label{ckansatz}
U=\sqrt{U_\pi} \, U_{K} \,  \sqrt{U_\pi} \ ,
\ee
where
\bea
U_K \ = \ \exp \left[ i\fr{\sqrt2}{{f_\pi}} \left( \begin{array}{cc}
                                                        0 & K \\
                                                        K^\dag & 0
                                                   \end{array}
                                           \right) \right] \ , \
                                           \ \ \
K \ = \ \left( \begin{array}{c}
                   K^+ \\
                   K^0
                \end{array}
                           \right),
\eea
and $U_\pi$ is the soliton background field written as
a direct extension to $SU(3)$ of the $SU(2)$ field $u_\pi$, i.e.,
\bea
U_\pi \ = \ \left ( \begin{array}{cc}
                       u_\pi & 0 \\
                       0 & 1
                    \end{array}
                               \right ) \ ,
\eea
with $u_\pi$ being the conventional hedgehog solution
$ u_\pi=\exp[i \vec \tau \cdot \hat{ r}F (r) ]$.

Assuming that the chiral symmetry breaking along the strangeness direction is
strong enough, the effective action is expanded up to the second 
order in the kaon field. 
The resulting lagrangian density can be written as the sum
of a pure $SU(2)$ term depending on $u_\pi$ only and an
effective lagrangian density describing the interaction between the soliton and the kaon
fields. The soliton profile $F(r)$ is obtained by minimizing the corresponding 
classical $SU(2)$ energy. On the other  hand, the kaon field satisfies an 
eigenvalue equation which describes its dynamics in the presence of the 
soliton background field.
In this picture, low--lying strange hyperons arise from the bound state solutions of this 
equation. In particular, the octet and decuplet hyperons are obtained by populating
the lowest kaon bound state which carries the quantum numbers $\Lambda=1/2,\ l=1$.
Here, $\Lambda$ is the grand--spin defined by the coupling of angular momentum and 
isospin and $l$ is the kaon angular momentum.  The splitting among hyperons with different
spin and/or isospin is given by the rotational corrections, which can be obtained
after introducing time-dependent rotations as $SU(2)$ collective coordinates. 
This approach has been shown to be successful in describing the hyperon
spectrum \cite{SNNR88} as well as other baryon properties such as the magnetic
moments \cite{OMRS91}. 

\section{The hyperon-nucleon interaction Lagrangian}
\label{pran}
In order to obtain the hyperon-nucleon potential we will approximate the
$B=2$ configuration by the product of two $B=1$ solutions, one centered at
$\vec x_1$ and the other one centered at $\vec x_2$. This is known as
the product ansatz and is a rather good approximation for studying the
medium and long distance behaviour of the potential. As well known, for short 
distances this approximation breaks down and  the exact solution
has a torus-like shape \cite{Ver87}. The kaon dynamics in the presence 
of the torus-like $B=2$ soliton configuration has been investigated in 
Ref.\cite{TSW94}. This is relevant for the study of strange exotics such as 
the $H$--particle. 

Within the product ansatz approximation the $B=2$ field is
written as
\be
U_{B=2}(\vec x ; \vec x_1,\vec x_2) = 
U_{B=1}(\vec x - \vec x_1) U_{B=1}(\vec x - \vec x_2)  \equiv U_1 U_2\ \ ,  \label{prodans}
\ee
where the indices $1,2$ indicate the dependence on  the coordinates of each 
individual soliton.

Substituting the ansatz (\ref{prodans}) and subtracting one--particle contributions, 
the resulting interaction lagrangian density coming from  the quadratic term ${\cal L}_2$ 
is 
\be \label{l2symm}
{\cal L}_2^{(int)} = \fr{\fpis}{4} \tr \l[ L_\mu^{1} R^\mu_2 +  
L_\mu^{2} R^\mu_1\r]  \ .
\ee
Here and in what follows we have performed an explicit symmetrization in the indices 
$1,2$ to ensure the invariance of the interaction under the exchange of the two particles. 

The quartic term leads to  
\bea
\label{l4int}
{\cal L}_4^{(int)} = 
\fr{1}{32 \eps^2} \Tr \Big[- 4 L_\mu^{1} L_\nu^{1} L^\mu_{1} R^\nu_{2} 
&+&  2 L_\mu^{1} L^\mu_{1} L_\nu^{1} R^\nu_{2} + 
2 L_\mu^{1} L_\nu^{1} L^\nu_{1} R^\mu_{2} \nn \\
- \ 4 L_\mu^{1} R_\nu^{2} R^\mu_{2} R^\nu_{2} &+&  
2 L_\mu^{1} R^\mu_{2} R_\nu^{2} R^\nu_{2} + 2 L_\mu^{1} R_\nu^{2} R^\nu_{2} R^\mu_{2} \nn \\
+ \ 4 L_\mu^{1} L_\nu^{1} R^\mu_{2} R^\nu_{2} &-&  
2 L_\mu^{1} L^\mu_{1} R_\nu^{2} R^\nu_{2} - 
2 L_\mu^{1} L_\nu^{2} R^\nu_{2} R^\mu_{2} \nn \\
+ \ 2 L_\mu^{1} R_\nu^{2} L^\mu_{1} R^\nu_{2} &-&  
 L_\mu^{1} R^\mu_{2} L_\nu^{1} R^\nu_{2} -  
L_\mu^{1} R_\nu^{2} L^\nu_{1} R^\mu_{2} \nn \\
+\ (1 \lra 2) \ \Big]  \ \ \; \ , & &  
\eea
while from the symmetry breaking term we obtain 
\bea
\label{lsbint1}
{\cal L}_{\rm SB}^{(int)} & = 
& { f_\pi^2 m_\pi^2 + 2 f_K^2 m_K^2 \over{24} }
 \Tr \left[ (U_1 - 1)(U_2 - 1) - 2 + h.c. \right] 
\nonumber \\
& &  
+ \sqrt{3}  { f_\pi^2 m_\pi^2 - f_K^2 m_K^2 \over{12} }
\Tr \left[ \lambda_8 \left( (U_1 - 1)(U_2 - 1) - 1 \right) +  
(1 \lra 2) + h.c. \right]  
\nonumber \\
& & - { f_K^2 - f_\pi^2\over{24} }
\Tr \left[ U_2 (1- \sqrt{3} \lambda_8) U_1 
\left(L_\mu^{1} - R_\mu^{2} \r) \left(L^\mu_{1} - R^\mu_{2} \r) 
+ (1 \lra 2) + h.c. 
\right] ,  
\eea
where $h.c.$ stands for hermitean conjugate and $(1 \lra 2)$ for the
exchange of the indices $1$ and $2$. 
The contribution from the Wess-Zumino term is  
\bea
{\cal L}^{(int)}_{\rm WZ} &=& 
-\fr{i N_c}{96 \pi^2} \Tr \l[ L_{1}^3 R_{2} - R_{2}^3 L_{1} - 
\fr{1}{2} L_{1}R_{2}L_{1}R_{2}  + (1 \lra 2) \r]
\eea
where the absence of Lorentz indices indicates that we have used the one form notation, i.e.,  
$L_{1}^3 R_{2} = \veps_{\mu\nu\alpha\beta} 
L^\mu_{1} L^\nu_{1} L^\alpha_{1} R^\beta_{2}$.

	The use of the ansatz (\ref{ckansatz}) for the individual chiral fields and
the subsequent expansion up to second order in the kaon components lead us to   
an interaction Lagrangian  that can be written as the sum of three different types of contributions. 
Namely, 
\bea 
{\cal L}^{(int)} &=& {\cal L}^{\pi d} + 
{\cal L}^{kd} + {\cal L}^{ke}  \ .
\eea 
A schematic representation of these interactions is shown in Fig.1.
Fig.1.a represents the $N_c^1$ direct--term ${\cal L}^{\pi d}$ which is a pure $SU(2)$ 
contribution, the bound kaon acting as a spectator, and 
Figs.1.b-c the two $N_c^0$ terms ${\cal L}^{kd}$ and ${\cal L}^{ke}$ 
where $K$ fields are present in direct and 
exchange interactions respectively. 
The ${\cal L}^{\pi d}$ and ${\cal L}^{kd}$ are both direct interactions in the
sense that the final particles are not exchanged with respect to the
initial state. The ${\cal L}^{kd}$ interaction corresponds to processes
where the kaon degrees of freedom are excited (and thus subleading in
$1 / N_c $ ), while the ${\cal L}^{ke}$ interaction involves the exchange of a kaon between the
particles and is called exchange contribution for short.

In general, each term in the effective action, Eq.(\ref{lag}), contributes
to the three different pieces ${\cal L}^{\pi d}$, ${\cal L}^{kd}$ and  
${\cal L}^{ke}$ in which we have split the interaction Lagrangian.
For the quadratic term of the action we get
\bea
\label{firstl2}
{\cal L}_2^{\pi d} &=& \fr{\fpic}{4} \Tr \l[ l_\mu^{1} r^{\mu}_2 + 
                                          l_\mu^{2} r^{\mu}_1 \r] \ , \\ 
\nn \\
{\cal L}_2^{kd} &=&  \fr{1}{4} \l( D^\mu K_2^\dag n_2^\dag l^{1}_\mu n_2 K_2
- K_2^\dag n_2^\dag l^{1}_\mu n_2 D^\mu K_2 \r) \nn \\
& & + \fr{1}{4} \l( D^\mu K_1^\dag n_1 r^{2}_\mu n_1^\dag K_1
- K_1^\dag n_1 r^{2}_\mu n_1^\dag D^\mu K_1 \r) \nn \\
 &  & - \fr{1}{8}  \l( K_2^\dag n_2^\dag r^{2\mu} l^{1}_\mu n_2 K_2
+ K_2^\dag n_2^\dag l^{1}_\mu r^{2\mu}  n_2  K_2 \r.\nn \\
& & \l.
\:\:\:\ \ \ \ \ +  \ K_1^\dag n_1  l^{1}_\mu r^{2\mu}  n_1^\dag K_1
+ K_1^\dag n_1 r^{2\mu}  l^{1}_\mu  n_1^\dag K_1 
\r) + (1 \lra 2) \ , \\
\nn \\
{\cal L}_2^{ke} &=& \frac{1}{2} \l( D_\mu K_1^\dag n_1 n_2 D^\mu K_2 +  
D_\mu K_2^\dag n_2^\dag n_1^\dag D^\mu K_1 \r) \nn \\
&  & +  \fr{1}{4} \l( D_\mu K_1^\dag n_1 r^{2\mu} n_2  K_2 + 
D_\mu K_2^\dag n_2^\dag l^{1\mu} n_1^\dag  K_1 \r) \nn \\
& & - \fr{1}{4} \l(  K_1^\dag n_1 l^{1\mu} n_2 D_\mu  K_2 +  
K_2^\dag n_2^\dag r^{2\mu} n_1^\dag  D_\mu K_1 \r) \nn \\
&  & - \fr{1}{8} \l(  K_1^\dag n_1 l^{1\mu} r^{2}_\mu n_2 K_2 + 
K_2^\dag n_2^\dag r^{2\mu} l^{1}_\mu n_1^\dag K_1 \r) + (1 \lra 2) \ , \\  \nn
\label{lastl2}  
\eea
where $n=\sqrt{u_\pi} $ and  we used the definitions
\bea
l_\mu &=& u_\pi^\dag \partial_\mu u_\pi \ \ ; \ r_\mu \ = \ u_\pi \partial_\mu u_\pi^\dag \ , \\
D_\mu  &=& \partial_\mu  + \fr{1}{2} ( n^\dag \partial_\mu n + n \partial_\mu n^\dag)  \ . 
\eea

In the case of the Wess-Zumino term the pure $SU(2)$ contribution ${\cal L}^{\pi d}_{\rm WZ}$
vanishes. The remaining two contributions are
\bea \label{wzdirect}
{\cal L}^{kd}_{\rm WZ}  &=& -\fr{\om N_c}{48 \pi^2 f_\pi^2}  
\Bigg[ K_2^\dag n_2^\dag l_1 r_2 n_2 DK_2 + DK_2^\dag n_2^\dag l_1 r_2 n_2 K_2 \nn \\
& & \ \ \ \ \ \ \ 
- \fr{1}{2} K_2^\dag n_2^\dag  \l( l_1^3 + l_1 r_2^2 - l_1 r_2 l_1 \r) n_2 K_2 \nn \\
& &  \ \ \ \ \ \ \ -  ( F_i \lra - F_i)  +  h.c.  \Bigg]  + (1 \lra 2) \ \ ,
\eea
and
\bea 
{\cal L}^{ke}_{\rm WZ}  &=& \fr{\om N_c}{48 \pi^2 f_\pi^2} 
\Bigg[ K_1^\dag n_1^\dag S n_2^\dag DK_2 + 
DK_1^\dag n_1^\dag S n_2^\dag K_2 \nn \\
& & \ \ \ \ \ \ \ \ + \fr{1}{2} K_1^\dag n_1^\dag \l( S l_2 - r_1 S \r) n_2^\dag K_2  
-( F_i \lra - F_i) \Bigg] + (1 \lra 2) \ ,
\eea
where  $F_i$ is the profile of the $i$th soliton  and  $ S = l_2^2 + r_1^2 - l_2 r_1 $. 
When performing $( F_i \lra - F_i \ ; \   i=1,2 ) $, 
the replacements $ l \lra r  , n \lra n^\dag  $ should be done.
Moreover, we have used that for bound antikaons $ \dot{K} = i \om K $,  with $\om > 0 $.

Similar expressions can be obtained for the Skyrme and symmetry breaking terms ${\cal L}_4$ 
and ${\cal L}_{\rm SB}$, respectively. Since their lengthy explicit forms are not 
particularly instructive we are not going to display them here.  

Next, the quantization of the two soliton system is performed using collective coordinates.  
We rotate the bound states, one  independently of the other
\bea
\label{colco}
& u_1 \ \rarr \ A_1 u_1 A_1^\dag & \ \ , \ \ u_2 \ \rarr \ A_2 u_2 A_2^\dag
\ , \nn \\
& K_1 \ \rarr \ A_1 K_1 & \ \ ,  \ \ K_2 \ \rarr \ A_2 K_2 \ , \
\eea
where $A_1$ and $A_2$ are  $SU(2)$ matrices. This dependence on the collective coordinates will be reexpressed in terms of the relative coordinate $ C = A_1^\dagger A_2 $. In order to obtain the physical 
particles we perform projections onto states with good spin and isospin quantum 
numbers. The corresponding general wavefunctions of the hyperons can be found
 e.g. in Ref.\cite{GRS92}.

Finally, to obtain the interaction potential we have to take matrix elements
of ${\cal L}^{(int)}$ between the relevant two--baryon wavefunctions and integrate 
out the center of mass coordinate $\vec R$. For the latter purpose it is convenient 
to express the individual positions of the particles in terms of $\vec R$ and their 
relative separation  $\vec r$ ,   
\bea
\vec x_1 &=& \vec R + {m_2\over{m_1+m_2}} \ \vec r \ , \nonumber \\
\vec x_2 &=& \vec R - {m_1\over{m_1+m_2}} \ \vec r \ ,
\eea
where $m_1$ and $m_2$ are the physical masses of the individual particles. 
We choose $\vec r$ to point in the $\hat z$ direction and perform the 
integration in $\vec x ' = \vec x - \vec R $. In this way we obtain 
\bea
V^{(int)}_{HN}(r) & = & - \int_0^\infty dR \; R^2 \int_{-1}^{1}  
d\eta \ \int_{0}^{2 \pi}  d\varphi  \  < \ {\cal L}^{(int)} \ >
\label{intdo}
\eea
where $\eta = \rh \cdot \hat R $.  
By performing the analytical integration over $\varphi$ one obtains an 
operator with a general structure that allows to identify the different components 
of the potential, like e.g. central component, spin-spin component, etc.
The remaining integrations over $R$ and $\eta$ are to be done numerically.

The formalism developed so far is valid for both $H = \Lambda, \Sigma$. 
Nevertheless, as several steps in this procedure imply long and involved calculations, in what follows we restrict ourselves to  the study of the $\Lambda N$ 
interaction potential. Since $\Lambda$ is an isoscalar particle the number of 
terms to be calculated is greatly reduced in this particular case.

\section{The $\Lambda N$ potential in the adiabatic approximation}
\label{angi}

We illustrate the general procedure to derive the interaction potential by 
considering some specific terms of the interaction Lagrangian. In this derivation
we neglect terms depending on the collective rotational
velocities (non-adiabatic terms). These terms would give rise to e.g. spin-orbit contributions and
are subleading by, at least, one order in $1/N_c$ with respect to the contributions 
considered here.
It should be noticed that, even within this approximation, the full calculation of all 
the terms contributing to the $\Lambda N$ potential is quite long.
To be confident of our results all the expressions were cross-checked by independent calculations, with the exception of the ${\cal L}_4$ kaonic contributions which could only be evaluated with the help of an algebraic computer code. 

\subsection{The direct contributions}

Let us start by considering a direct interaction  of the ${\cal L}^{\pi d}$-type.
One sees that there is no ${\cal L}_2$ contribution of this kind
to the $\Lambda N$ potential. The reason is that, after the introduction of the  
collective coordinates, eq.(\ref{firstl2}) contains the expression 
\bea
\label{cdc}
C^\dag l_1^j C &=& l_{1a}^j C^\dag \sigma_a C = l_{1a}^j R_{ab}(C) \sigma_b \ ,
\eea 
with  $R_{ab}$ a rotation operator defined by eq.(\ref{rij}). Because of eq.(\ref{pfd2}) this leads
to a vanishing $\Lambda N$ matrix element. 
The most important contribution comes from ${\cal L}_4$ , 
which is responsible for the central repulsion. 
It can be easily obtained by replacing $L$ and $R$ from eq.(\ref{l4int}) by 
their $SU(2)$ counterparts, since the kaon acts here as a spectator. 
After taking matrix elements and replacing in eq.(\ref{intdo}), we
obtain 
\bea
V^{\pi d}_C(r) &=& \fr{2 \pi}{3 \eps^2} \int_0^\infty dR \; R^2 \int_{-1}^{1}  d\eta \ \Bigg[
\l(F_1' F_2'\r)^2 + \l(F_1' \fr{s_2}{x_2}\r)^2 + \l(F_2' \fr{s_1}{x_1}\r)^2 \nn \\
& & \ \ \ \ \ \ \ \ \ \ \ \ + 3 \l(\fr{s_1 s_2}{x_1 x_2 }\r)^2 - 
\l(F_1'^2 - \l(\fr{s_1}{x_1}\r)^2 \r) \l( F_2'^2 - \l(\fr{s_2}{x_2}\r)^2 \r) 
( \hat x_1 \cdot \hat x_2 )^2 \Bigg] 
\ ,
\eea
where $s_i = \sin F_i$ , $ c_i = \cos F_i$ and $F_i' =dF_i/dx_i$. 

As an example of the treatment of the ${\cal L}^{kd}$--type terms we take the
contribution from the Wess--Zumino term, eq.(\ref{wzdirect}). Again, 
after the introduction of collective coordinates, all terms containing 
expression (\ref{cdc}) vanish. In this case, however, there are still two additional terms. One of these terms gives simply
\bea
\label{l3}
C^\dag l_1^3 C = - 6 F_1' \fr{s^2_1}{x^2_1} \ ,
\eea
since $l^3$ is an isoscalar proportional to the $SU(2)$ contribution to the baryon density. 
This  is the only non-vanishing contribution. 
The other term contains the expression 
\bea
C^\dag l_1 C r_2 C^\dag l_1 C &=& \veps_{ijk} l^i_{1a} r^j_{2b} l^k_{1c} C^\dag \sigma_a C 
\sigma_b C^\dag \sigma_c C \nn \\
&=& \veps_{ijk} l^i_{1a} r^j_{2b} l^k_{1c} \sigma_d \sigma_b \sigma_f R_{ad}(C) R_{cf}(C) \  
\eea 
and although the corresponding collective matrix element is non-zero, the total matrix 
element vanishes due to the antisymmetry of $\veps$--tensor. 
Therefore, from eq.(\ref{l3}) we finally obtain
\bea
\int  d\varphi <\Lambda_2' N_1' |{\cal L}^{kd}_{\rm WZ} |\Lambda_2 N_1>  &=& 
 - \fr{ \omega N_c}{8 \pi^2 f_\pi^2} k_2^2 F_1' \fr{s_1^2}{x_1^2} \ {\cal O}_C  \ .
\eea

It should be mentioned that the terms containing more than two $C^\dagger,C$ pairs 
give non-zero contributions to the direct ${\cal L}_4$ 
interactions since no $\veps$--tensor is present in that case.

\subsection{The exchange contributions}

To illustrate the calculation of the exchange contributions we consider
the first two terms in the symmetrized form of ${\cal L}_2^{ke}$. 
After introducing collective coordinates and neglecting
non-adiabatic terms they reduce to
\bea
\fr{1}{2}  D_\mu K_1^\dag n_1 C n_2 D^\mu K_2 + (F_i \lra - F_i ) + h.c.
\eea
Using Eq.(\ref{pfe1}) the relevant matrix element reads
\bea 
\label{exl2}
<\Lam'_1 N'_2 | \fr{1}{2}  D_\mu K_1^\dag n_1 C n_2 D^\mu K_2 |\Lam_2 N_1> & = & 
 \fr{1}{4}  \delta_{I_3^N , I_3^{N'}} < J^{\Lam'}_3 | D_\mu K^\dag n  | J^N_3 >_{1} 
< J^{N'}_3 | n D^\mu K |J^{\Lam}_3  >_{2}  \nn \\
& & 
\eea
The individual  matrix elements can be calculated using a projection theorem given in Ref.\cite{GRS92}. Given the explicit form of the hegdehog ansatz we obtain
\bea \label{expl1}
n_2 D^0 K_2 \!\! & \!\! = \!\! & \!\! \fr{\om k_2}{2 \sqrt{\pi}} 
\l( \ss_2 - i \cs_2 \vec \sigma_2 \cdot \hat x_2 \r) \ , \nn \\
n_2 D^a K_2 \!\! & \!\! = \!\! & \!\! \fr{1}{2 \sqrt{\pi}} \l[ i k_2' \ss_2 \ \hat x_2^a + 
\l( \cs_2 k_2' - \cs_2^3 \fr{k_2}{x_2} \r) \vec \sigma_2 \cdot \hat x_2 \ 
\hat x_2^a + 
\cs_2^3 \fr{k_2}{x_2} \ \sigma_2^a + \cs^2_2  \fr{k_2}{x_2}
\ss_2 \ (\hat x_2 \times \vec \sigma_2 )^a \r]
\eea
where $k'$ stands for the radial derivative of the kaon wavefunction. 
Moreover, we use 
the short-hand notation
\bea
\ss = \sin \fr{F}{2} \   & , & \cs = \cos \fr{F}{2} \ . 
\eea
Similar expressions are obtained for the operator $D_\mu K_1^\dagger n_1$. 
In this way, one obtains the explicit form of the matrix element, eq.(\ref{exl2}). 
Next, we integrate out the center of mass coordinate. 
At this stage it is convenient to define the operators 
\bea 
\hat {\cal O}_C  & = & \left(I\right)_{\Lambda'N} \left(I\right)_{N'\Lambda} \ ,
\nonumber \\ \  
\hat {\cal O}_S  & = & \left(\vec \sigma_1\right)_{\Lambda'N} 
\cdot \left(\vec \sigma_2\right)_{N'\Lambda} \ , 
\nonumber \\  
\hat {\cal O}_T  & = & 3 \ \rh \cdot \left(\vec \sigma_1\right)_{\Lambda'N} 
 \  \rh \cdot \left(\vec \sigma_2\right)_{N'\Lambda}  -  
\left(\vec \sigma_1\right)_{\Lambda'N} \cdot \left(\vec \sigma_2\right)_{N'\Lambda} 
\eea
and make use of the relation
\bea
\int_{0}^{2 \pi}  d\varphi  \  \vec \sig_1 \cdot \hat x_i \ \vec \sig_2 \cdot \hat x_j  
&=& \fr{2 \pi}{3} \l( x_{ij} \hat {\cal O}_S + G_{ij} \hat {\cal  O}_T  \r) 
\eea 
with $x_{ij} = \hat x_i \cdot \hat x_j$ and
\bea
G_{ij} = \fr{3 R^2}{2 x_i x_j } (\eta^2 -1) +  x_{ij} \ . 
\eea
Using the expressions given above we obtain the complete result    
\bea
& & \int d\varphi <\Lambda'_1 N'_2 | \left[ 
\fr{1}{2}  D_\mu K_1^\dag n_1 C n_2 D^\mu K_2 + (F_i \lra - F_i ) + h.c. 
\right] | \Lambda_2 N_1 > \  = \nn \\  
&=&  \quad  - \fr{1}{12} \delta_{I_3^N , I_3^{N'}}  \Bigg\{ 
\om^2 k_1 k_2 \l( 3 \ss_1 \ss_2 \hat {\cal O}_C - \cs_1 \cs_2 x_{12} \hat {\cal O}_S - 
\cs_1 \cs_2 G_{12} \hat {\cal  O}_T \r) \nn \\
& &  \:\:\:\ \ \ \ \ \ \  + \hat {\cal O}_C \l( -3 k_1' k_2' \ss_1 \ss_2 x_{12} \r)  \nn \\
& &  \:\:\:\ \ \ \ \ \ \  + \hat {\cal O}_S \Bigg[ \fr{k_1 k_2}{x_1 x_2} \cs_1^2 \cs_2^2 \l( 
\cs_1 \cs_2 ( 1 + x_{12}^2 ) - 2 \ss_1 \ss_2 x_{12} \r) \nn \\
& & \:\:\:\ \ \ \ \ \ \ \ \ \ \ \ \ \ \   +  \fr{k_1 k_2'}{x_1} \cs_1^3 \cs_2 (1-x_{12}^2) +  
\fr{k_1' k_2}{x_2} \cs_1 \cs_2^3 (1-x_{12}^2) \nn \\
& &  \:\:\:\ \ \ \ \ \ \ \ \ \ \ \ \ \ \  +  k_1' k_2' \cs_1 \cs_2 x_{12}^2 \Bigg] \nn \\
& & \:\:\:\ \ \ \ \ \ \   + \hat {\cal O}_T  \Bigg[ - \fr{k_1 k_2}{x_1 x_2} \cs_1^2 \cs_2^2 
\Big( \cs_1 \cs_2 ( G_{11} + G_{22} - G_{12} x_{12}) - \ss_1 \ss_2 G_{12} \Big) \nn \\
& &   \:\:\:\ \ \ \ \ \ \ \ \ \ \ \ \ \ \  +  \fr{k_1 k_2'}{x_1} \cs_1^3 \cs_2 
(G_{22} - G_{12} x_{12}) +  \fr{k_1' k_2}{x_2} \cs_1 \cs_2^3 ( G_{11} - G_{12} x_{12} ) \nn \\
& & \:\:\:\ \ \ \ \ \ \ \ \ \ \ \ \ \ \  +  k_1' k_2' \cs_1 \cs_2 G_{12} x_{12} \Bigg] \:\:\:\: 
\Bigg\} \ .
\eea

In order to recover the operator structure of the potential as given in eq.(\ref{pot}) 
we still have  to perform a Fierz rearrangement and write the operators 
$\hat{\cal O}_C , \hat{\cal O}_S , \hat{\cal O}_T  $ in terms of
the operators  ${\cal O}_C , {\cal O}_S , {\cal O}_T  $ appearing in
such equation. We obtain  
\bea
\alpha \ \hat {\cal O}_C + \beta \ \hat {\cal O}_S + \gamma \ \hat {\cal O}_T 
&=& \fr{1}{2} \l( \alpha + 3 \beta \r) {\cal O}_C
+ \fr{1}{2} \l( \alpha - \beta \r) {\cal O}_S + \gamma  \ {\cal O}_T  \ , 
\eea
where $\alpha, \beta$ and $\gamma$ are arbitrary functions depending only
on the relative separation $r$.   
All the other exchange contributions to the interaction potential can be treated in exactly 
the same way.

\section{Numerical results and discussion}

In the numerical calculations we used the physical values for the different mesonic parameters 
appearing in the Lagrangian, that is 
$m_\pi = 138 \ MeV ,  \ m_K = 495 \ MeV, \ f_\pi = 93 \ MeV , \ f_K/f_\pi = 1.22$ and 
take $\eps = 4.26$ in order to fit the mass difference between the 
nucleon and the $\Delta$. 
With these values the hyperon excitation spectrum is rather well described. 
On the other hand, the absolute baryon masses come out too high by about 800 MeV. 
This is a generic problem of the topological soliton models that can be fixed
by properly taking into account the quantum corrections to the soliton 
mass \cite{Mou93}. Recently, this has been explicitly shown in  the case
of the bound state soliton model \cite{SW98}.

The individual contributions of the different direct
and exchange terms to the $\Lambda N$ potential 
are shown in Fig.2 and Fig.3 , respectively.
As a general feature we see that the pionic contributions
are much larger (in absolute value) than the kaonic ones
as expected from $N_c$-counting. 
We also notice that in the present 
scheme the direct terms only contribute 
to the central potential. In particular, although there is
an attractive symmetry breaking contribution to $V_C^{\pi d}$,
such part is completely dominated by the repulsive contribution 
coming from the quartic term. As already mentioned there are no
${\cal L}_2$- and ${\cal L}_{\rm WZ}$-contributions of this type.
In the case of the direct kaonic part the quartic and WZ contributions are 
attractive and similar in magnitude. As seen in Fig.3 all the different 
terms in the Lagrangian contribute to the exchange potentials. 
All these contributions are attractive except for those
coming from the symmetry breaking term ${\cal L}_{\rm SB}$ which are, in any case,
quite small.

Our total predictions for the central, spin-spin and tensor
components of the $\Lambda N$ potential are presented in Fig.4.
As anticipated from the discussion above, we observe that the spin--spin and tensor 
interactions are supressed by an order of magnitude with respect to the central
interaction which turns out to be repulsive at any distance. Noting that
$V_C$ is dominated by the pionic contributions it is  clear that such
a behaviour is very much related with the well known problem of the missing
central attraction at intermediate distances in the $SU(2)$ Skyrme model.
As reviewed in Ref.\cite{EK96} many mechanisms have been proposed to solve
this problem.  Whatever such solution could be, our
$SU(3)$ calculations show that the inclusion of strangeness degrees of
freedom are not likely to spoil it since the central kaonic 
contributions are attractive. 

The sign of our predicted spin-spin contribution implies that there will be 
more attraction in the $^3S_1$ channel than in the $^1S_0$.
The empirical information about the sign of the spin-spin interaction is
somewhat
unclear. From the existing $\Lambda p$ scattering data it is very
difficult to draw
a definite conclusion. In fact, various versions of the OBE model that fit
the scattering data 
equally well lead to rather different predictions for the $^3S_1$ and $^1S_0$ 
scattering lengths  (see e.g. Ref.\cite{nsc97}).
On the other hand, the hypernuclei data tend to favor a repulsive $\Lambda
N$ spin-spin
interaction although again the question is not completely settled. The most
clear
indication comes from the $^4_\Lambda H$ and $^4_\Lambda He$ doublet
states. However,
the analysis of these states depends on non-trivial four-body calculations.
There
is also some empirical information from other hypernuclei like e.g.
$^{11}_\Lambda B$.
There, however, the situation is even more complicated because of the role
played by
spin-orbit interactions. Forthcoming experiments on both hyperon-proton 
scattering \cite{Ier98}
and hypernuclear $\gamma$ spectroscopy \cite{Tam98}
are expected to provide critical tests on this issue.    
>From the point of view of the Skyrme model our results are consistent with
those obtained
in the previous $SU(3)$ collective coordinates  calculations \cite{KE90,SSG95} in the
absence of
channel couplings. One might argue that since in our model the pion
exchanges are
taken into account beyond the OBE a good deal of the mixing with the
$\Sigma N$ channel
is taken into account. However, our approach still allows
for non-vanishing off-diagonal $N\Lambda-N\Sigma$ terms. There are indications
that when such mixing with rotational excited configurations is included
the sign
of the spin-spin interaction in the Skyrme model might be reversed \cite{KE90}.

Finally, for $r>1.2 \ fm$ our prediction
for the tensor component of the potential agrees well with the OBE models.
For smaller distances there are large discrepances between OBE model
D and F. For example, at $r\approx .9 \ fm$ one has $V_T^{OBE-D} \approx  - 14 \ MeV$ 
while $V_T^{OBE-F} \approx  + 9 \ MeV$. In such region our results favor
 those of model D.

The dashed lines in Fig.4 represent the results of the collective approach \cite{SSG95}
to the $SU(3)$ Skyrme model. We see that the magnitude and sign of the 
potentials are similar to those of the bound state approach. However, 
the situation is different for the behaviour at large
distances. Although in both cases $V_C$ decays basically in the same
way, the range of the $V_S$ and $V_T$ in the CCA is much longer.
This is not difficult to understand since in the CCA the only meson
that determines the fall-off of the radial functions is the pion meson. On 
the other hand, in the BSA the spin-spin and tensor components of the $\Lambda N$ 
potential are given by the kaonic components. The corresponding 
range is, therefore, associated with the kaon mass which is several 
times larger than the pion mass. In the central potential
these differences do not appear because of the dominance of the
pionic ${\cal L}_4$-contribution already discussed.
It should be mentioned that in all the cases the behaviour of the 
potentials at large distances obtained in the BSA is in good agreement
with the results of the OBE models.         

\section{Conclusions}

We have investigated the hyperon-nucleon two-body interaction in the
framework of the bound state approach to the $SU(3)$ Skyrme model. We would like to 
stress the fact that the Skyrme model approach incorporates chiral symmetry and 
the large $N_c$ expansion in an elegant way. Moreover, 
by relating the physics of sectors with different baryonic numbers it gives 
parameter free predictions for the present calculation, in contrast 
with more phenomenological approaches. 
 Our studies
have been based on the product ansatz for the $B=2$ soliton field which is adequate
for the medium and large separation distances discussed in this work. 
We have found that there are three 
classes of contributions within the adiabatic approximation used here. One type
corresponds to order $N^1_c$ purely $SU(2)$ contributions in which kaons act as
simply spectators. The other two are of order $N_c^0$ and correspond to direct
and exchange kaonic interactions. Although the formalism we have followed is 
suitable for any diagonal or off-diagonal hyperon-nucleon potential we have concentrated 
on the diagonal $\Lambda N$ interaction. There, important simplifications appear in the 
expressions for the potentials which still happen to be lengthy and cumbersome.
We have found that the  central potential is repulsive at any distance. This is
strongly related to the missing intermediate range central attraction of the 
$NN$ potential as calculated in the $SU(2)$ Skyrme model. Within our scheme, any of the 
suggested solutions of this problem is expected to bring in some attraction
also in the $\Lambda N$ case. 
Generally speaking our results
are very similar to those of the $SU(3)$
collective coordinate approach to the 
Skyrme model \cite{SSG95}. An exception to this is the range of the spin-spin
and tensor interactions. For these quantities the values obtained in the 
present calculation seem to be more realistic. 
Finally, there are  some  indications that the coupling to vibrations could give
the missing central attraction while the coupling to rotationally excited states 
may change the sign of the spin-spin interaction \cite{KE90}. The formalism 
developed in the present work provides a general framework for future
investigations of such issues within the bound state soliton model.

\section*{Acknowledgements}

This work was partially supported by a grant of the Fundaci\'on Antorchas, Argentina 
and FAPERGS and FINEP, Brasil. N.N.S. also acknowledges a grant of the ANPCYT, Argentina. 
The authors would like to express their gratitude to the late Prof. J.M. Eisenberg for stimulating 
discussions. C.L.S. and N.N.S. would also like to thank Prof. A. Gal for valuable conversations 
regarding the phenomenology of $\Lambda N$ potentials. C.L.S. thanks M.Kruczenski for initial 
guidance on the use of algebraic computer codes and the CNPq-Brasil for financial support 
through a CLAF fellowship.

\section*{Appendix }

\renewcommand{\theequation}{A.\arabic{equation}}
\setcounter{equation}{0}

In this appendix we present a set of formulae which are useful for calculating  
the matrix elements of the collective lagrangian operators. 

The rotation operator written in the cartesian basis 
\bea \label{rij}
R_{ab}(C) &=& \fr{1}{2} {\Tr} \l[ \sigma_a C \sigma_b C^\dag \r] 
\eea  
can be expressed in terms of the spherical tensor 
$D_{\alpha\beta}(C)$ in the following way
\bea
\label{dax}
R_{ab} &=& \eh_\alpha \cdot \hat e_a \ \eh^*_\beta \cdot \hat e_b \ D_{\alpha\beta}, 
\eea
where $\eh_\alpha$, with $\alpha = +1, 0, -1$, are the usual spherical unit vectors
and $\hat e_a$ the cartesian ones.

Using Eq.(\ref{dax}) it is not difficult to show that
\bea
\l(
D^{(1/2)}_{m n}\r)^* D^{(1/2)}_{m' n'} &=& \fr{1}{2} \l( \delta_{m m'} 
\delta_{n n'} \ + \  <m'|\sigma^a |m> R_{ab} <n|\sigma^b |n'> \r) \ .
\eea
The evaluation of the matrix elements amounts to an integration over $SU(2)$.
For products of $R_{ab}$ we have
\bea
\fr{1}{2 \pi^2} \int dC \ R_{ab} R_{cd} &=&
\fr{1}{3} \delta_{ac} \delta_{bd} \ ,  \\
\fr{1}{2 \pi^2} \int dC \ R_{ab} R_{cd} R_{ef} &=&
\fr{1}{6} \eps_{ace} \eps_{bdf} \ .
\eea
Using explicit forms of the $\Lambda$ and $N$ wavefunctions
\bea
|\Lambda> &=& \fr{1}{\sqrt{2} \pi}\ |\frac{1}{2} J_3^{\Lambda}> \ , \nonumber \\ 
| N > &=& \frac{i}{\pi} (-1)^{ \frac{1}{2} + I_3^N} D^{(1/2)}_{-I_3^N,J_3^N} \ ,
\eea
together with the expressions above,  the relevant collective matrix elements can 
be easily calculated.
For the direct terms we have 
\bea \label{pfd1}
< \Lam'_2 N'_1 |  \Lam_2 N_1> &=&  
\delta_{I_3^N , I_3^{N'}} \delta_{J_3^N , J_3^{N'}} \delta_{I_3^\Lam , I_3^{\Lam'}} \ \ \ , \\
\label{pfd2}
< \Lam'_2 N'_1| R_{ab}(C) | \Lam_2 N_1 > &=& 0 \ , \\
 < \Lam'_2 N'_1|R_{ab}(C) R_{cd}(C) |\Lam_2 N_1 > &=&
\fr{1}{3} \delta_{ac} \delta_{bd} < \Lam'_2 N'_1|  \Lam_2 N_1>
\label{pfd3}
\eea
and for the exchange terms
{\small
\bea \label{pfe1}
<\Lam'_1 N'_2 |{\cal O}^\dag_\mu (K_1) C_{\mu\nu} {\cal O}_\nu (K_2) |\Lam_2 N_1  > 
 \!\!\!&\!\!\!= \!\!\!& \!\!\! 
{\delta_{I_3^N , I_3^{N'}}\over2} < J^{\Lam'}_3 |{\cal O}^\dag (K_1) | J^N_3 > 
< J^{N'}_3 |{\cal O} (K_2) |J^{\Lam}_3  > \ , \nonumber \\
 \\
<\Lam'_1 N'_2 |{\cal O}^\dag_\mu (K_1) C_{\mu\nu} {\cal O}_\nu (K_2) R_{ab}(C) |\Lam_2 N_1  > 
\!\!\!&\!\!\!= \!\!\!& \!\!\!  
{\delta_{I_3^N , I_3^{N'}}\over6} 
< J^{\Lam'}_3 |{\cal O}^\dag (K_1) \sigma_a| J^N_3 >
< J^{N'}_3 |\sigma_b{\cal O} (K_2) |J^{\Lam}_3  > \nonumber \\
\label{pfe2}
\eea
}


\pagebreak

\vspace*{4.cm}
\centerline{ { \psfig{figure=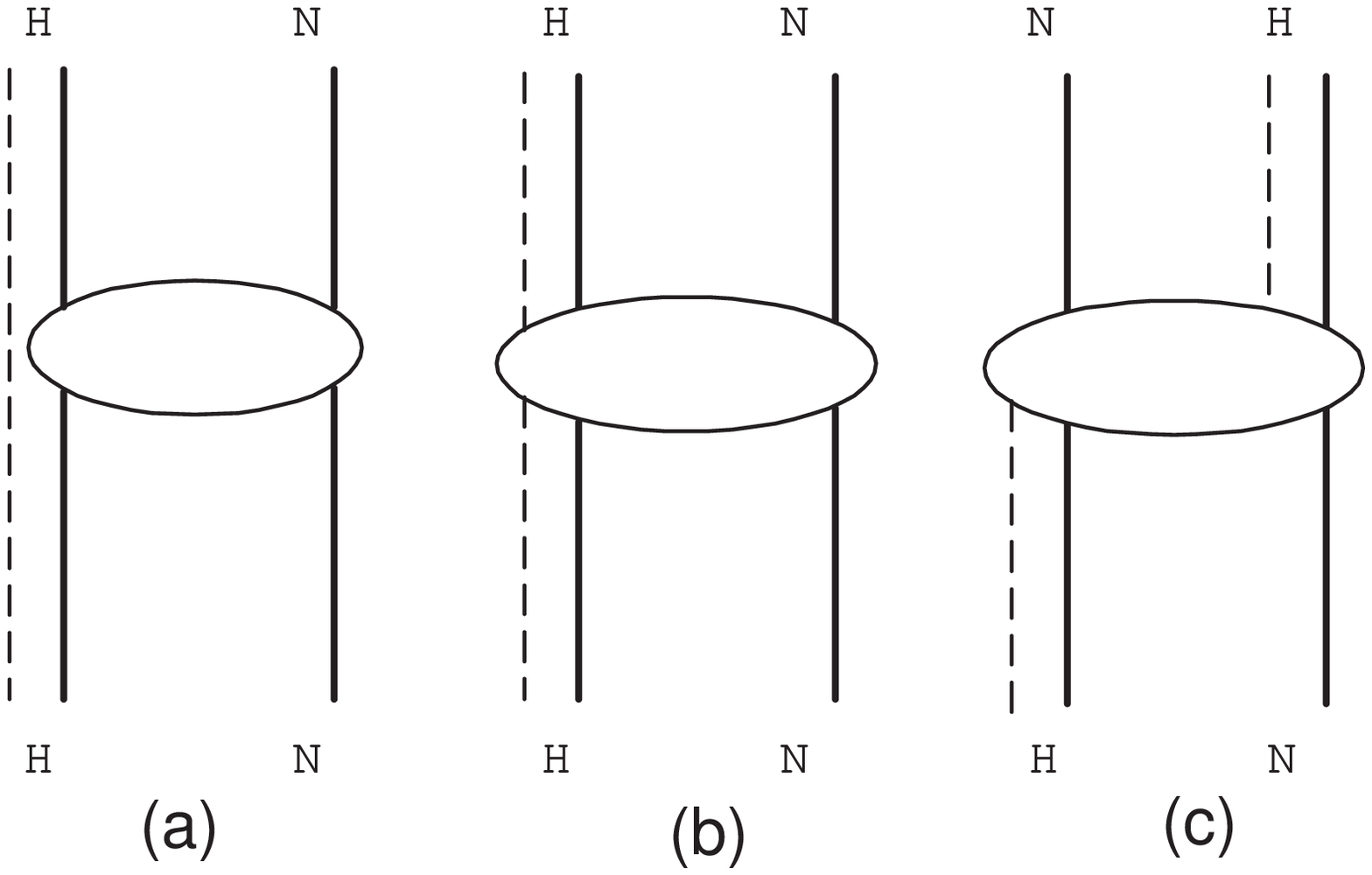,width=12.cm,height=6.cm,angle=0} } }
\vskip .5cm
\centerline{ \parbox{14cm} 
{{\scriptsize {Fig.1 -- Schematic representation of the different types
of contributions to the hyperon-nucleon potential. (a) Direct ``pionic"
contributions ${\cal L}^{\pi d}$, (b) direct ``kaonic" contributions
${\cal L}^{kd}$ and (c) exchange ``kaonic" contributions ${\cal L}^{k e}$ .} }} }

\pagebreak

\vskip .1cm
\centerline{ { \psfig{figure=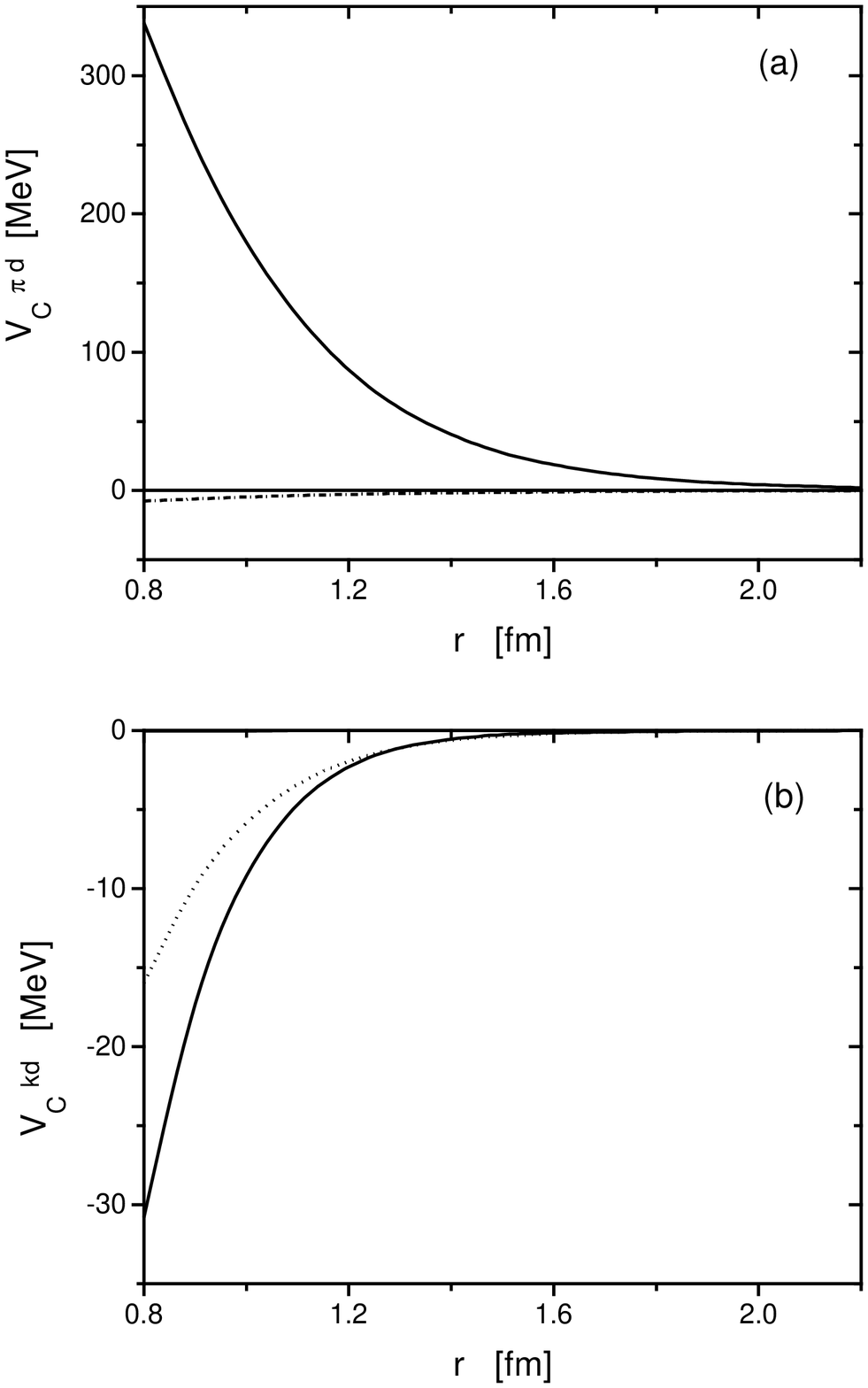,width=15.cm,height=20.cm,angle=0} } }
\vskip .5cm
\centerline{ \parbox{15cm} {{\scriptsize 
{Fig.2 -- 
(a) Central contributions to the $\Lambda N$ potential coming from  
the direct ``pionic" terms, ${\cal L}^{\pi d}$. 
(b) Central contributions to the $\Lambda N$ 
potential coming from direct ``kaonic" terms ${\cal L}^{kd}$. The full line 
represents the contributions from ${\cal L}_4$, the dotted line those from ${\cal L}_{\rm WZ}$ 
and the dashed-dotted line those from ${\cal L}_{\rm SB}$. } }} }

\pagebreak

\vskip .1cm
\centerline{ { \psfig{figure=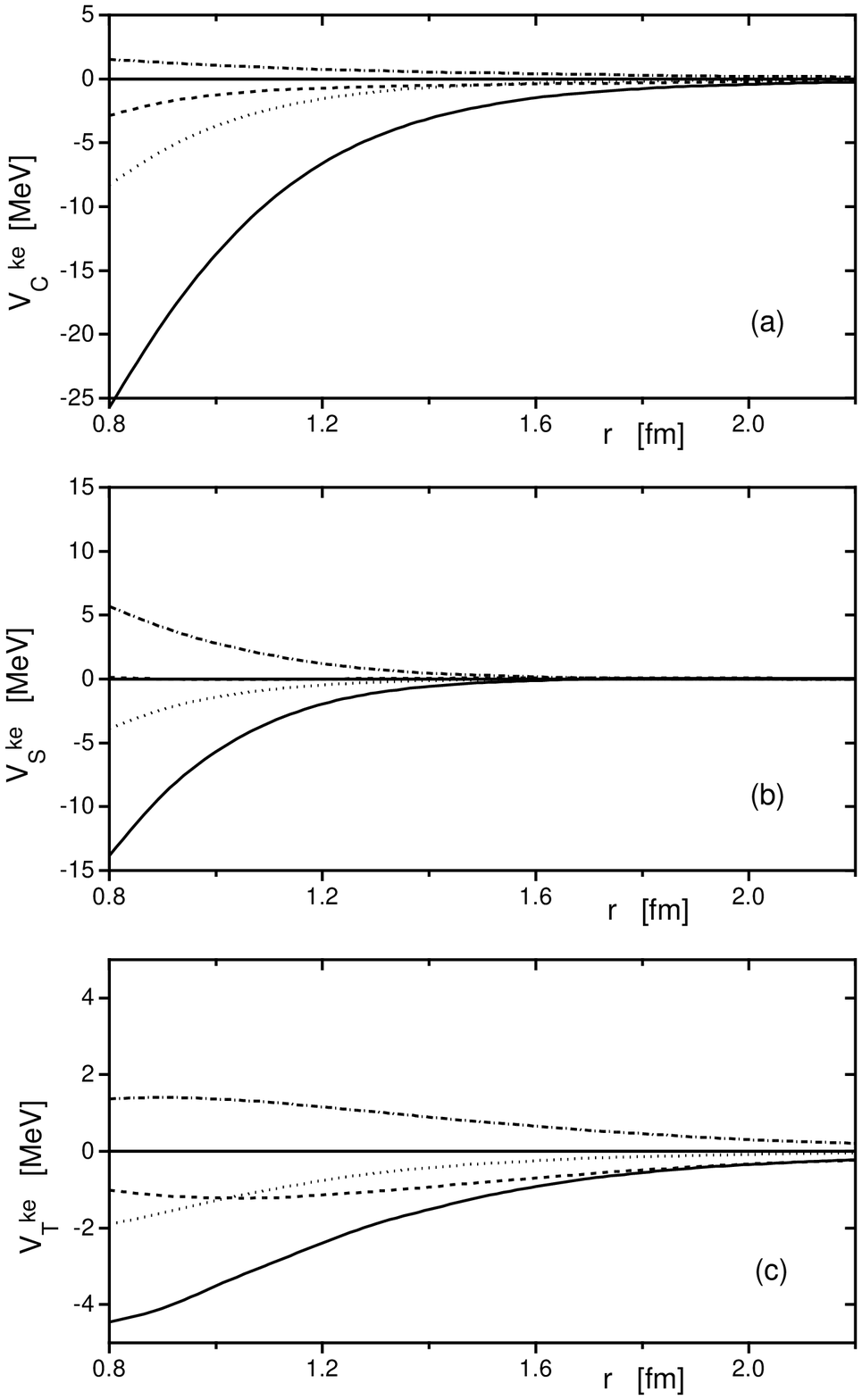,width=15.cm,height=20.cm,angle=0} } }
\vskip .5cm
\centerline{ \parbox{15cm} {{\scriptsize 
{Fig.3 -- Contributions from  
the exchange ``kaonic" terms ${\cal L}^{ke}$ to: 
(a) Central component of the $\Lambda N$ potential, 
(b) Spin-spin component of the $\Lambda N$ potential,
(c) Tensor component of  the $\Lambda N$ 
potential. In the three panels the dashed line represents the contributions
from ${\cal L}_2$, the full line those from ${\cal L}_4$, the dotted line those from 
${\cal L}_{\rm WZ}$ and the dashed-dotted line those from ${\cal L}_{\rm SB}$. } }} }

\pagebreak

\vskip .1cm
\centerline{ { \psfig{figure=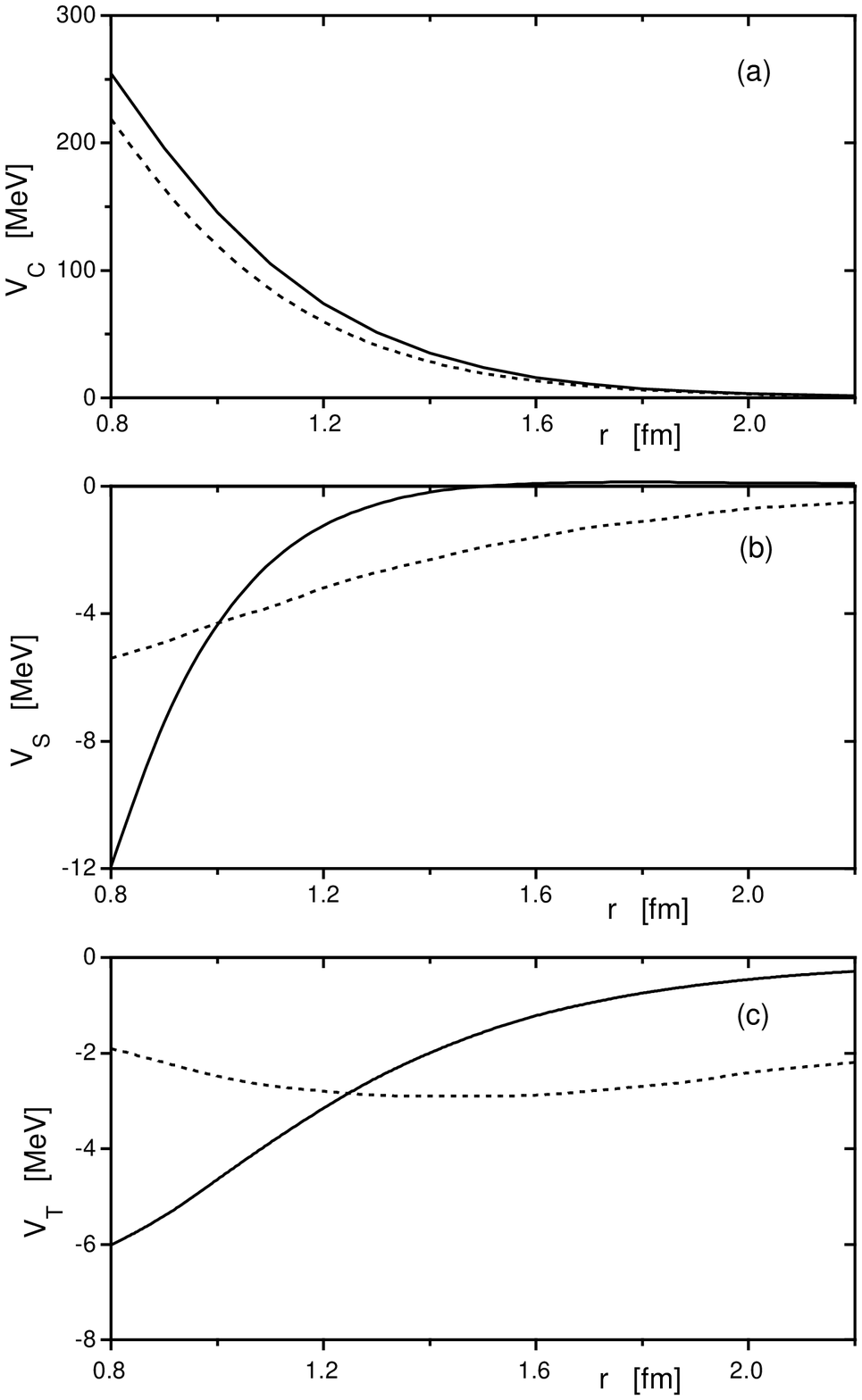,width=15.cm,height=20.cm,angle=0} } }
\vskip .5cm
\centerline{ \parbox{15cm} {{\scriptsize 
{Fig.4 -- Components of the $\Lambda N$ potential as defined in Eq.(\ref{pot}): 
(a) central component $V_C$, 
(b) spin-spin component $V_S$ and 
(c) tensor component $V_T$. In the three panels the full line represents the results
of the present calculation and the dashed line those of the CCA as given in Ref.\cite{SSG95}} 
}} }

\end{document}